\documentclass[%
reprint,%
 amssymb, amsmath,%
 aip,apl,%
]{revtex4-1}

\usepackage{docs}%
\usepackage{bm}%
\usepackage[colorlinks=true,linkcolor=blue, citecolor=red]{hyperref}%
\expandafter\ifx\csname package@font\endcsname\relax\else
 \expandafter\expandafter
 \expandafter\usepackage
 \expandafter\expandafter
 \expandafter{\csname package@font\endcsname}%
\fi
\usepackage{graphicx}

\begin{document}
\title{ Formation of pentagonal atomic chains in BCC Fe nanowires}

\author{G. Sainath}%

\email{sg@igcar.gov.in}
\affiliation{Deformation and Damage Modelling Section, Metallurgy and Materials Group, 
Indira Gandhi Center for Atomic Research, HBNI, Kalpakkam, Tamilnadu -603102, India}%

\author{B.K. Choudhary}

\email{bkc@igcar.gov.in}

\affiliation{Deformation and Damage Modelling Section, Metallurgy and Materials Group, 
Indira Gandhi Center for Atomic Research, HBNI, Kalpakkam, Tamilnadu -603102, India}

\begin{abstract}

For the first time, we report the formation of pentagonal atomic chains during tensile deformation of ultra 
thin BCC Fe nanowires. Extensive molecular dynamics simulations have been performed on $<$100$>$/\{110\} BCC 
Fe nanowires with different cross section width varying from 0.404 to 3.634 nm at temperatures ranging from 
10 to 900 K. The results indicate that above certain temperature, long and stable pentagonal atomic chains 
form in BCC Fe nanowires with cross section width less than 2.83 nm. The temperature, above which the 
pentagonal chains form, increases with increase in nanowire size. The pentagonal chains have been observed 
to be highly stable over large plastic strains and contribute to high ductility in Fe nanowires.\\ 


\end{abstract}

\maketitle

\section{Introduction}

In recent years, metallic nanowires have attracted a major attention for research due to their
unique properties and potential applications in future nano/micro electro-mechanical systems.
The nanowires suspended between two electrodes are found to exhibit the electron transport
properties similar to that of a quantum wire which shows a quantized conductance, and it has
gained importance in fundamental physics and in electronic device technology as a quantized
conductance atomic switch \cite{Atomic-switch}. Under tensile loading, if the cross-section 
area of nanowire reduces to that of a few-atom thickness prior to failure, it exhibits 
conductance in multiples of $G_0 = 2e^2/h$, where e is the electron charge and h is the 
Planck’s constant \cite{Nature-1D}. In view of this correlation between the structure and the 
conductance, it is important to understand the formation and the stability of the long atomic 
chains during deformation of metallic nanowires. Further, the 1-D structures can also act as 
ideal systems to understand the low-dimensional physics.

In the past, the formation of the linear, pentagonal and other multi-shell atomic chains
have been observed during the deformation of many FCC nanowires and the formation of all
these atomic chains is associated with the quantized conductance \cite{Nature-1D,Nano-Lett,
PRB-Galvao,Exp-Galvao,PRL-Zigzag,Kondo-Science}. The long linear atomic chains have been 
observed experimentally during the stretching of Au and Cu nanowires \cite{Nature-1D,Nano-Lett,
PRB-Galvao}. The linear atomic chains display conductance close to $G_0$. Using transmission 
electron microscopy, Gonzalez et al. \cite{Exp-Galvao} observed the formation of pentagonal
chains during the stretching of Cu nanowires which exhibited the conductance of 4.5$G_0$.
During tension deformation, the Au nanowires have been found to exhibit a zigzag shape,
which remains stable even if the load is relieved \cite{PRL-Zigzag}. Similarly, Kondo and 
Takayanagi \cite{Kondo-Science} synthesized the gold nanowires with diameter as small as 0.6 
nm having length of 6 nm. Using high-resolution electron microscopy, it has been found that 
these nanowires have a multi-shell structure composed of coaxial tubes  \cite{Kondo-Science}. 
Each tube consists of helical atom rows coiled round the wire axis. The structures such as 
linear atomic chain (monatomic, diatomic), pentagonal, two-shell, three-shell and helical 
multi-shell structures have also been predicted theoretically in ultra thin Al and Pb metallic 
nanowires \cite{Noncrystal} and also in ZnO and Si semiconductor nanowires using molecular 
dynamics simulations \cite{Semi-Cond1,Semi-Cond2}. In Al and Pb metallic nanowires, it has 
been postulated that the competition between optimal internal packing and minimal surface 
energy is the main driving force behind these morphological structures at small size 
\cite{Noncrystal}. However, out of all these structures, the formation of pentagonal atomic 
chains is one of the fascinating observations as it exhibits a five-fold symmetry with respect 
to the nanowire axis and does not correspond to any of the 14 Bravais lattices. Generally, 
the pentagonal structure is not observed in 2-D or 3-D structures as it is incompatible with 
the translational symmetry. However, as mentioned above, the strong evidence of pentagonal
structures have been found experimentally in 1-D structures, such as ultra-thin Cu nanowires 
\cite{Exp-Galvao}. In agreement with experimental observations, many atomistic simulation 
studies have also shown the formation of pentagonal atomic chains in Cu \cite{Sutrakar1,Sutrakar2,
MSMSE}, Al \cite{MSMSE}, Ni \cite{MSMSE,Ni}, and Au \cite{Park-PRB,Park1}. The pentagonal 
structures have also been observed in HCP Mg nanowires during the atomistic simulations 
studies \cite{HCP-Mg1,HCP-Mg2}. The evidence of pentagonal structures has been found during 
the stretching of BCC Na nanowires using first-principles molecular dynamics simulations 
\cite{Na}[19]. Other than Na, the pentagonal structure has not been reported in any other
metallic BCC nanowires either by experiments or atomistic simulations. For the first time 
in this paper, we report the formation of long pentagonal atomic chains in ultra thin BCC 
Fe using molecular dynamics (MD) simulations. Further, the formation mechanism and the
range of nanowire size and temperatures over which pentagonal atomic chains form are presented.

\section{Molecular dynamics simulations}

Molecular dynamics (MD) simulations have been performed using LAMMPS package \cite{LAMMPS} employing an 
embedded atom method (EAM) potential for BCC Fe given by Mendelev and co-workers \cite{Mendelev}. BCC Fe 
nanowires oriented in $<$100$>$ axial direction with \{110\} as side surfaces were considered in this 
study. Simulations have been performed on nanowires with cross section width (d) ranging from 0.404 to 
3.634 nm. In all nanowires, the length (l) was twice the cross section width. Periodic boundary conditions 
were chosen along the nanowire length direction, while the other directions were kept free in order to 
mimic an infinitely long nanowire. After the initial construction of nanowire, energy minimization was 
performed by conjugate gradient method to obtain a stable structure. Velocity verlet algorithm was used 
to integrate the equations of motion with a time step of 5 femto seconds. Finally, the model system was 
thermally equilibrated to a required temperature in canonical ensemble with a Nose-Hoover thermostat. 
For each nanowire cross section width, the simulations have been performed at different temperatures 
ranging from 10 to 900 K. Following thermal equilibration, the tensile deformation was performed at a 
constant strain rate of  $1\times$ $10^8$ $s^-1$ along the axis of the nanowire. The strain rate 
considered during deformation is significantly higher than the typical experimental strain rates. This 
is due to the inherent timescale limitations from MD simulations. Here, it must be noted that the strain 
rate also influences the formation of linear and pentagonal atomic chains \cite{Semi-Cond1,Sutrakar2}. 
For each size and temperature, five independent simulations have been carried out to make statistically 
meaningful conclusions. The average stress was calculated from the Virial expression \cite{Virial}. The
visualization of atomic configurations was performed using AtomEye package \cite{AtomEye} with coordination 
number and common neighbour analysis (CNA). In solid-state systems, the CNA pattern is useful for 
obtaining local crystal structure around an atom \cite{CNA1,CNA2}. In the present analysis, this method 
is used distinguish the atoms in crystalline, non-crystalline and pentagonal or icosahedral environment.

\section{Results and Discussion}

\begin{figure}[h]
\centering
\includegraphics[width=7.5cm]{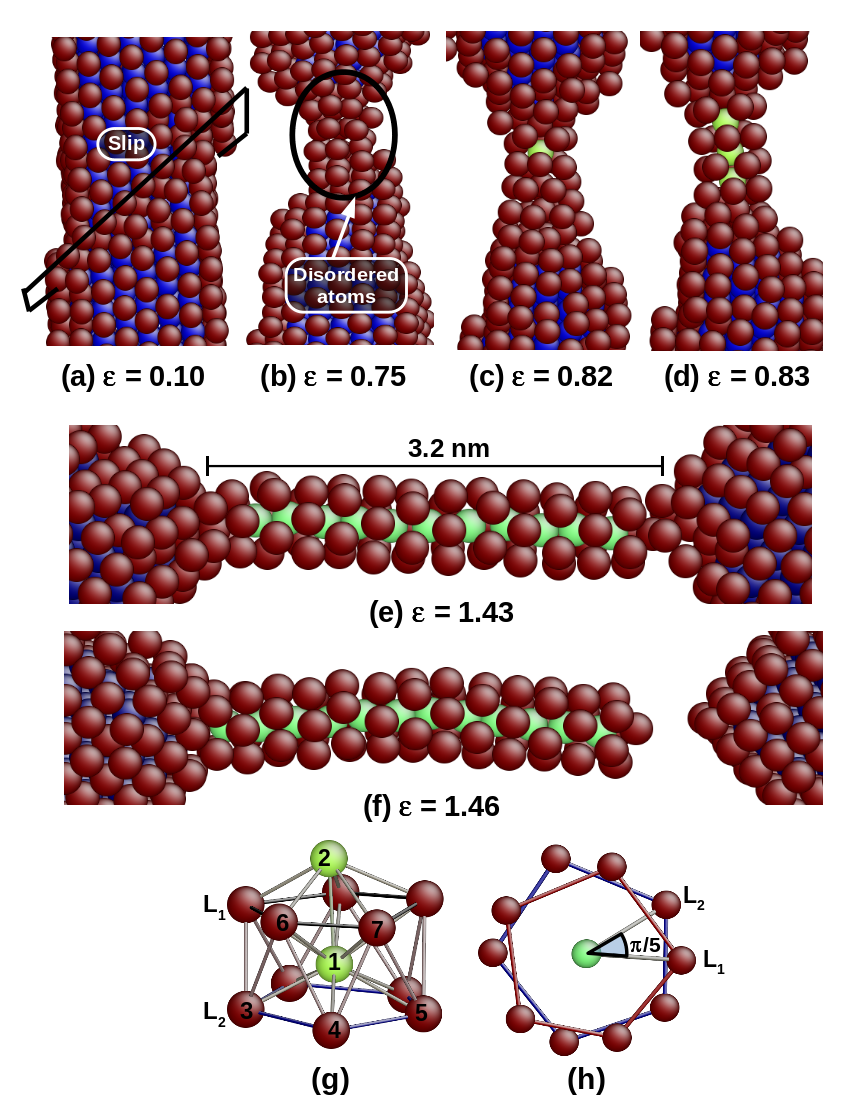}
\caption {The formation of pentagonal atomic chain in the necking region of BCC Fe nanowire with d = 1.615 nm during 
tensile deformation at 600 K. The atoms are coloured according to CNA. The blue colour indicates the atoms in BCC
structure, the red colour indicates the atoms in disordered structure including surfaces and light green colour 
indicates the atoms in pentagonal or icosahedral symmetry.}
\label{Pentagon}
\end{figure}

Molecular dynamics simulations have been carried out on BCC Fe nanowires with cross section width varying 
from 0.404 to 3.634 nm at different temperatures ranging from 10-900 K. The atomic snapshots demonstrating 
the formation of pentagonal atomic chains before failure during tensile deformation of BCC Fe nanowire are 
typically shown for d = 1.615 nm at 600 K in Fig. \ref{Pentagon}. Initially, the nanowire undergoes yielding 
followed by an extensive plastic deformation dominated by the slip of 1/21/2$<$111$>$ full dislocations 
(Fig. \ref{Pentagon}a). With increasing strain, the plastic deformation occurs on multiple slip systems 
and leads to necking with disordered or non-crystalline atomic structure shown in Fig. \ref{Pentagon}(b). 
When the cross section of the neck region is close to a few atomic spacings, some of the disordered atoms
rearrange themselves and forms pentagonal unit cell shown in Fig. \ref{Pentagon}(c). Following this, the
five atom rings are added successively to the structure unit by unit at the expense of disordered atoms in 
the necking region and this leads to increase in the length of pentagonal atomic chain (Fig. \ref{Pentagon}d).
This continuous process occurs until failure resulting in the formation of long pentagonal atomic chain in 
the necking region (Figs. \ref{Pentagon}(e-f)). The maximum attained length of the pentagonal atomic chain 
has been obtained to be 3.2 nm consisting of 13 pentagonal rings. The formation of long pentagonal chains 
indicates the stability at high strains and this contributes to large plastic deformation and high ductility 
in small size BCC Fe nanowires. In order to obtain detailed understanding of structural aspects of pentagonal 
atomic chain, a single unit cell of the long chain is presented in Fig. \ref{Pentagon}(g). It can be seen 
that the pentagonal chain consist of a central atom having a coordination of 10 atoms enclosed by two 
pentagonal rings $L_1$ and $L_2$. The long atomic chain has 1-5-1-5 stacking sequence with successive 
pentagonal rings $L_1$ and $L_2$ rotated by $\pi/5$ (Fig. \ref{Pentagon}h). This structure is known as 
staggered pentagon and has 12 atoms in each unit cell. The stability of staggered pentagonal structure in 
Fe is confirmed by Sen et al. \cite{Sen} using first principle calculations. The relevant inter atomic 
distances and the bond angles in the pentagonal unit cells for small size (d = 0.807 nm) and large size 
(d = 1.615 nm) Fe nanowire at 600 K are listed in Table \ref{Table}. A good agreement between the inter-atomic 
distances and bond angles obtained in the present study and those calculated by Sen et al. \cite{Sen} for 
free pentagonal nanowires using first principle calculations can be clearly seen in Table \ref{Table}.

\begin{table*}[t]
\caption{ Comparison of inter-atomic distances and the bond angles in pentagonal unit cells 
for small (d = 0.807 nm) and large (d = 1.615 nm) size nanowires at 600 K with those reported based on 
first principles calculations. $d_{ij}$ indicates the distance between the atoms i and j in Figure 
\ref{Pentagon}(g) and $\angle ijk$ indicates the bond angle between the i, j and k atoms.}
\centering
\scalebox{1.0}
{
\begin{tabular}{llllllllll} \\
\hline
& $d_{12}$ & $d_{13}$ & $d_{34}$ & $d_{46}$ & $\angle 314$ & $\angle315$ & $\angle316$ & $\angle317$ & $\angle364$ \\
 \hline
Small (d = 0.87 nm) & 2.40 & 2.50 & 2.62 & 2.72 & 62.17 & 110.38 & 65.52 & 111.62 & 57.29 \\
Large size (d = 1.42 nm) &  2.37 & 2.56 & 2.75 & 2.74 & 65.15 & 116.4 & 66.16 & 119.8 & 58.74\\
First principles calculations \cite{Sen} & 2.16 & 2.31 & 2.39 & 2.51 & 62.4 & 114.2 & 65.8 & 117.5 & 57.00 \\

\hline
\end{tabular}
}
\label{Table} 
\end{table*}

\begin{figure}[h]
\centering
\includegraphics[width=7.5cm]{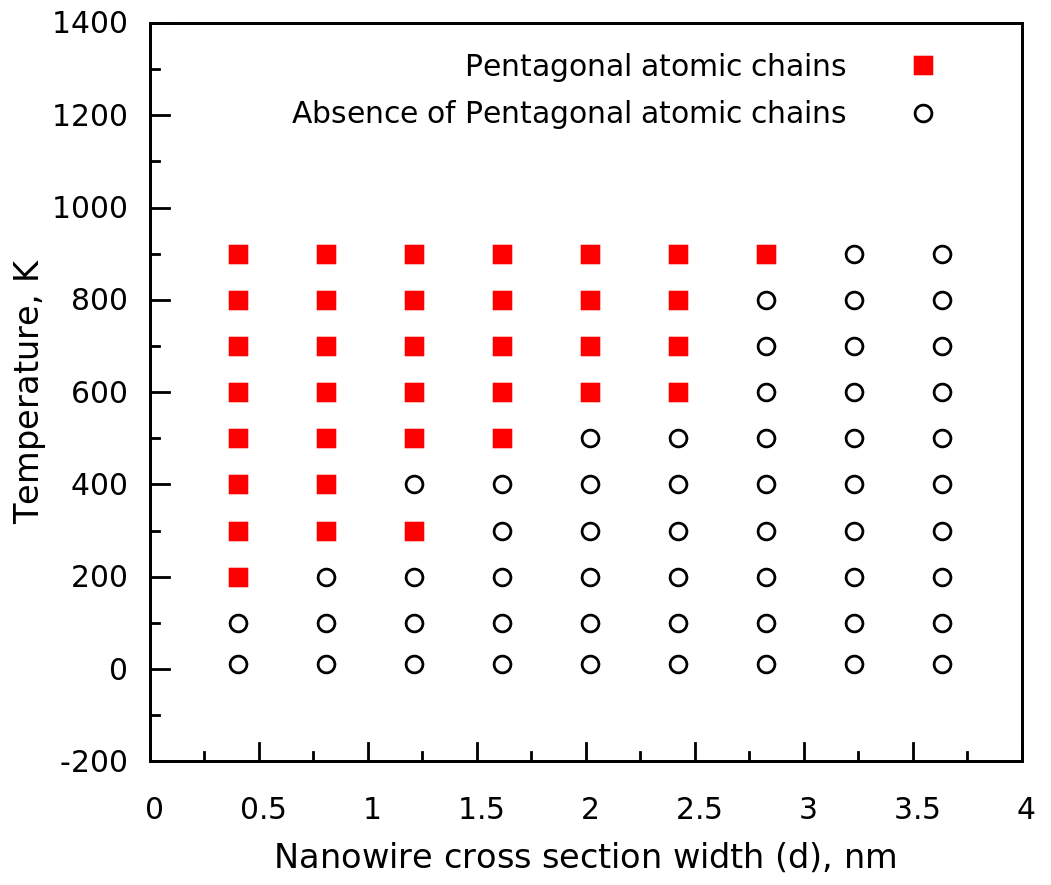}
\caption { Temperature vs. nanowire size plot showing the regions of size and temperature over which pentagonal 
atomic chains (PAC) are formed.}
\label{Size-Temp}
\end{figure}

The formation of pentagonal atomic chains as a function of the combination of nanowire size and temperature 
is shown in Fig. \ref{Size-Temp}. It can be clearly seen that the formation of pentagonal atomic chains is 
promoted by decrease in nanowire size and increase in temperature. The range of nanowire size over which 
pentagonal atomic chains are formed increases with increase in temperature. In fact, for each nanowire size, 
a transition temperature exists above which the pentagonal atomic chains are formed. The transition temperature 
increases linearly with increase in nanowire size. It is important to point out that the pentagonal atomic 
chains are not formed at low temperatures of 10 and 100 K as well as nanowire size larger than 2.83 nm. The 
observed formation of pentagonal atomic chains promoted by small nanowire size and higher temperatures is in 
agreement with those reported for FCC and HCP nanowires \cite{Sutrakar1,Sutrakar2,MSMSE,Ni,Park-PRB,Park1,
HCP-Mg1,HCP-Mg2}. In addition to above, it has been shown that the crystallographic orientation and strain 
rate influence the formation of pentagonal atomic chains during tensile deformation in FCC nanowires 
\cite{Sutrakar2,MSMSE,Ni}. The combined effect of temperature and strain rate on the formation of linear 
atomic chains in ZnO nanowires has been explained based on the Arrhenius model \cite{Semi-Cond1}. Further, 
the observation of transition temperature as a function of size in BCC Fe nanowires appears to be significant, 
since there is no such transition temperature established yet in either FCC \cite{Sutrakar1,Sutrakar2,MSMSE,Ni,
Park-PRB,Park1} or HCP nanowires \cite{HCP-Mg1,HCP-Mg2}. These results indicates that at proper size and 
temperature, the BCC Fe nanowires can also exhibit long pentagonal atomic chains during the deformation.

\begin{figure}[h]
\centering
\includegraphics[width=7.5cm]{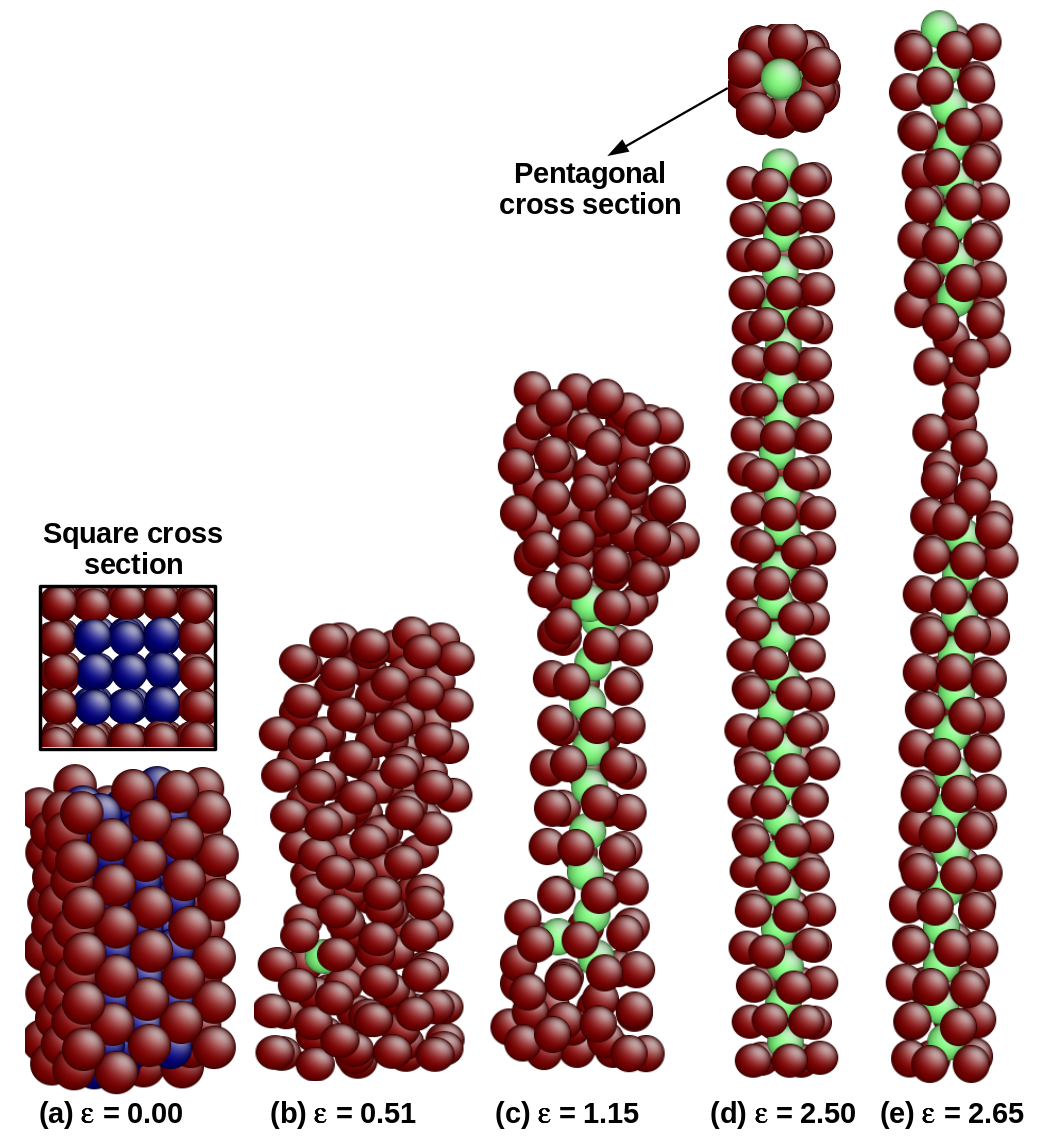}
\caption {The complete transformation of BCC Fe nanowire with d = 0.807 nm into long pentagonal nanowire at 600 K. 
(a) Initial BCC structure, (b) the disordered structure along with nucleation of pentagonal atomic chain, (c) 
growth of pentagonal atomic chain (d) completely transformed into pentagonal atomic chains and (e) onset of necking 
and failure. The atoms are coloured according to CNA. The blue colour indicates the atoms in BCC structure, the red 
colour indicates the atoms in disordered structure including surfaces and light green colour indicates the atoms 
in pentagonal or icosahedral symmetry.}
\label{Small-Penta}
\end{figure}

The pentagonal atomic chains are observed for the wide range of nanowire sizes and temperatures as shown in 
Fig. \ref{Size-Temp}. However, there has been significant difference between the pentagonal chains formed 
in the small and comparatively large size nanowires. In large size nanowires, the formation of pentagonal 
atomic chains remains limited to necking regions only as observed for nanowire width, d = 1.615 nm in 
Fig. \ref{Pentagon}. In case of small size nanowires with d = 0.404 and 0.807 nm, complete transformation 
of full nanowire into pentagonal atomic chains is observed at relatively higher temperatures. This is typically
shown for nanowire width d = 0.807 nm at 600 K in Figs. \ref{Small-Penta}(a-e). The progressive plastic
deformation resulting in the complete transformation into non-crystalline state and the nucleation of 
pentagonal atomic chains can be seen in Figs. \ref{Small-Penta}(b) and \ref{Small-Penta}(c), respectively. 
The complete transformation of full nanowire into pentagonal atomic chains and failure at large strain are 
presented in Figs. \ref{Small-Penta}(d) and \ref{Small-Penta}(e), respectively. In small size nanowires, 
the process of pentagonal structure formation remains same as that of the large nanowires. However, the plastic 
deformation in small size nanowires transforms the complete nanowire into a non-crystalline state (Figs. 
\ref{Small-Penta}(a-b)), unlike only the neck area in large size nanowires. The complete transformation into 
pentagonal atomic chains has been observed in nanowires with cross-section width d = 0.404 and 0.807 nm at 
temperatures above 400 and 500 K, respectively (Fig. \ref{Size-Temp}). The transformation into non-crystalline 
structure (Fig. \ref{Small-Penta}b) can be understood in terms of internal atomic rearrangements to maximize 
the overall atomic coordination leading to an increase in cohesive energy \cite{MSMSE}. Following the disordered 
state, the pentagonal unit cell nucleates out of this to minimize the surface energy (Fig. \ref{Small-Penta}c) 
and grows all along the nanowire leading to long and stable pentagonal nanowire (Fig. \ref{Small-Penta}d). In 
small size nanowires (d = 0.807 nm), the long pentagonal atomic chains consisted of as many as 25 pentagonal 
rings compared to 13 rings in large size nanowire with d = 1.615 nm at 600 K. The observed long pentagonal 
atomic chains remain stable during the deformation and contribute to large failure strains as much as 265\% in 
small size nanowires.

\begin{figure}[h]
\centering
\includegraphics[width=8cm]{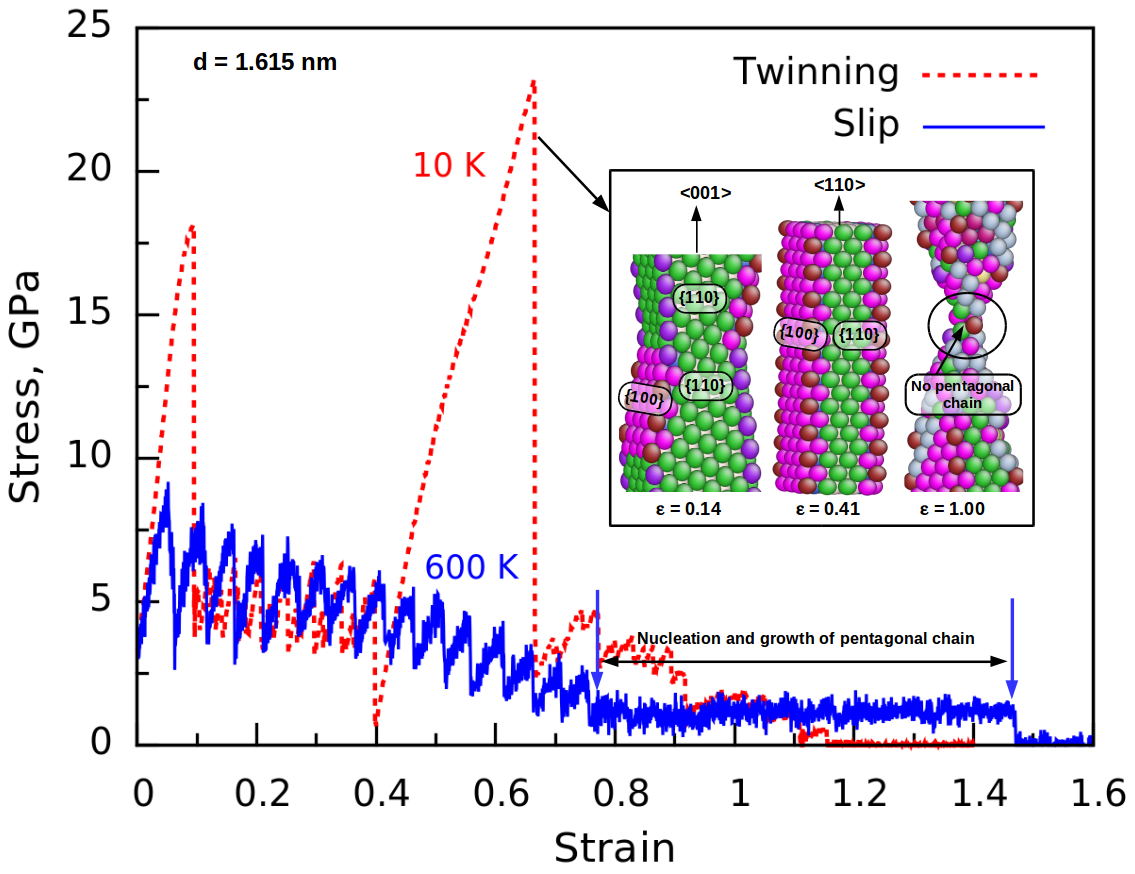}
\caption { Typical stress-strain behaviour exhibiting occurrence of pentagonal chains at 600 K and its absence at 
10 K for BCC Fe nanowires with d = 1.615 nm are shown by full and broken lines respectively. The nucleation and 
growth of pentagonal atomic chains at 600 K are marked in the Figure. The nanowires which do not exhibit pentagonal 
structure and deform by twinning followed by reorientation are shown in the inset at 10 K.}
\label{stress-strain}
\end{figure}

Typical stress-strain behaviour of BCC Fe nanowires exhibiting the formation of pentagonal chain at 600 K and 
in absence of pentagonal chain at 10 K is shown in Fig. \ref{stress-strain} for nanowire with d = 1.615 nm. 
In case of nanowire showing the formation of pentagonal atomic chains, large flow stress oscillations in terms 
of sharp peaks and drops due to continuous plastic deformation by dislocation slip can be seen up a strain of 
0.77. Here, each flow stress peak and drop corresponds to dislocation nucleation, propagation and annihilation. 
Once neck forms in the nanowire, the nucleation of pentagonal atomic structure and its growth along the nanowire 
axis leads to constant plateau as observed in the strain range 0.77-1.47. In general, the formation of pentagonal 
atomic chains facilitates high ductility in BCC Fe nanowires. On the other hand, the nanowires that do not show 
the formation of pentagonal structures exhibit different stress-strain behaviour in terms of flow stress oscillations 
and the second elastic peak. Unlike deformation by slip in nanowires showing pentagonal chains, the plastic 
deformation by twinning leading to complete reorientation from initial $<$100$>$/\{110\} to $<$110$>$/\{100\} 
orientation has been observed in nanowires that do not show the formation of pentagonal structures. The large 
second peak in stress-strain behaviour at 10 K is due to the elastic deformation of the reoriented nanowire. With 
increasing strain, this reoriented nanowire undergoes deformation by slip, but this has not lead to diffused 
necking and as a result, the pentagonal atomic chains were not observed. Detailed studies on $<$100$>$ BCC Fe 
nanowires have shown that the deformation by twinning followed by reorientation occurs at 10 K \cite{Sainath1,
Sainath2,Sainath3}. This result shows that the orientation nanowire also influence the formation of pentagonal 
atomic chains in BCC Fe nanowires.

\section{Conclusions}
We have performed extensive molecular dynamics simulations on the tensile deformation of ultra thin BCC Fe nanowires. 
The results indicated that long pentagonal atomic chains are formed in the diffused necking region of small size 
$<$100$>$/\{110\} BCC Fe nanowires. The complete transformation to the pentagonal structures has been observed in 
nanowires with d = 0.404 and 0.807 nm at higher temperatures. The temperature above which the pentagonal atomic 
chains are formed increases with increase in the size of nanowire, and above a critical size of 2.83 nm, the pentagonal 
atomic chains are not observed at any temperature. BCC Fe nanowires exhibiting pentagonal structure undergo deformation 
by slip mechanism. The nanowires which do not exhibit pentagonal structure deform by deformation twinning mechanism 
and undergo reorientation from the initial $<$100$>$/\{110\} orientation to $<$110$>$/\{100\}\{110\} orientation. \\

\end{document}